\documentclass[]{revtex4}
\usepackage{amsmath}
\begin{document}

\def\a{\alpha}
\def\b{\beta}
\def\e{\epsilon}
\def\p{\partial}
\def\m{\mu}
\def\n{\nu}
\def\t{\tau}
\def\s{\sigma}
\def\g{\gamma}
\def\r{\rho}
\def\half{\frac{1}{2}}
\def\hatt{{\hat t}}
\def\hatx{{\hat x}}
\def\hatp{{\hat p}}
\def\hatX{{\hat X}}
\def\hatY{{\hat Y}}
\def\hatP{{\hat P}}
\def\hatth{{\hat \theta}}
\def\hatta{{\hat \tau}}
\def\hatrh{{\hat \rho}}
\def\hatva{{\hat \varphi}}
\def\barx{{\bar x}}
\def\bary{{\bar y}}
\def\barz{{\bar z}}
\def\p{\partial}
\def\nn{\nonumber}
\def\cb{{\cal B}}
\def\2pap{2\pi\alpha^\prime}
\def\wideA{\widehat{A}}
\def\wideF{\widehat{F}}
\def\beq{\begin{eqnarray}}
\def\eeq{\end{eqnarray}}
\def\2pap{2\pi\a^\prime}
\def\xp{x^\prime}
\def\xpp{x^{\prime\prime}}
\def\xppp{x^{\prime\prime\prime}}
\def\barxp{{\bar x}^\prime}
\def\barxpp{{\bar x}^{\prime\prime}}
\def\barxppp{{\bar x}^{\prime\prime\prime}}
\def\barchi{{\bar \chi}}
\def\bpsi{{\bar \psi}}
\def\barg{{\bar g}}
\def\barz{{\bar z}}
\def\bareta{{\bar \eta}}
\def\ta{{\tilde a}}
\def\tb{{\tilde b}}
\def\tc{{\tilde c}}
\def\tpsi{\tilde{\psi}}
\def\tPsi{\tilde{\Psi}}
\def\tal{{\tilde \alpha}}
\def\tbe{{\tilde \beta}}
\def\barth{{\bar \theta}}
\def\bareta{{\bar \eta}}
\def\barom{{\bar \omega}}

\title[Short Title]{The Final Fate of the Rolling Tachyon}

\author{Taejin Lee}
\email{taejin@kangwon.ac.kr}

\affiliation{Department of Physics, Kangwon National University,
Chuncheon 200-701, Korea}


\affiliation{Pacific Institute for Theoretical Physics,
Department of Physics and Astronomy, University of British Columbia,
6224 Agricultural Road Vancouver, B.C. V6T 1Z1, Canada}

\date{\today}

\begin{abstract}
We propose an alternative interpretation of the boundary state for the rolling
tachyon, which may depict the time evolution of unstable D-branes in string 
theory. Splitting the string variable in the temporal direction into the 
classical part, which we may call ``time" and the quantum one, 
we observe the time dependent behaviour of the boundary. 
Using the fermion representation of the rolling tachyon boundary state, 
we show that the boundary state correctly describes the time-dependent 
decay process of the unstable D-brane into a S-brane at the classical level.
\end{abstract}

\pacs{04.60.Ds, 11.25.-w, 11.25.Sq}
\maketitle

\section{Introduction}

The time evolution of unstable states has been a perennial 
research subject in various branches of theoretical physics.
The most recent problem which appears on the stage is the fate of unstable
D-branes in string theory. Since A. Sen brought up this problem \cite{sen1}, 
called rolling tachyon, it has been one of the most important research themes 
in string theory 
\cite{senreview,lambert,sen022,mukhopa,sen023,sen0305,sen031,sen0306,sen0308,sen04,
larsen02,gibbons,okuda,kim,hlee,rey,rey2,taka,doug,arefeva,demasure,gutperle03,larsen,
constable,schomerus,nagami,foto,coletti,forini}. 
It was asserted that the unstable D-brane decays as the open 
string tachyons condensate \cite{harvey00,kuta,noncomm,okuyama,andreev00,aruty,
david,chalmers,lee0105,lee01052,lee0109,andreev,hotta,uesugi,ellwood,hashimoto} 
on the D-brane and it may be described by an exact 
time-dependent classical solution of string theory. 

The rolling tachyon can be described in the sigma model approach 
by introducing an exactly marginal boundary term to the world-sheet string action,
\beq \label{tachyonaction}
S = -\frac{1}{4\pi}
\int_M d\tau d\sigma \p X^0\cdot \p X^0 + \oint_{\p M} d\s
\left(Ae^{X^0}+Be^{-X^0}\right) +\ldots,
\eeq 
where  the second term describes the boundary interaction corresponding 
to the open string tachyon field with $A$ and $B$ constants. 
We abbreviate the string action for the spatial string coordinates $X^i$, 
$i= 1,\dots,25$ and use units where $\a^\prime =1$. Note the negative signature of 
the kinetic term for $X^0$ which indicates that it is the time coordinate. 
Choosing $A=g/2$ and $B=0$, we can study an unstable D-brane which decays into a 
more stable configuration. The tachyon profile with this simplest choice is called 
half-S-brane \cite{lambert}. The quantum theory of this sigma model is well 
defined if we take a Wick rotation to Euclidean time, $X^0 
\rightarrow iX^0$. After Wick rotation we get a conformal theory 
for a non-compact boson with a periodic boundary potential. It is 
well known that the conformal field theory of this kind can be 
described in terms of the boundary state, which can be given as an 
exact sum of Ishibashi states of the $SU(2)$ current algebra. 
Sen observed that this boundary state can be used to depict the 
time evolution of the unstable D-branes. Since then, various aspects of 
the time dependent process driven by the tachyon instability have
been explored by numerous authors. However, our understanding the 
final fate of the unstable D-brane and the exact dynamics of the 
rolling tachyon is not yet complete. The exact time dependent description
is only available for the first few lower levels.
If we expand the boundary state in the bosonic oscillator basis, 
we get
\beq \label{expansion}
|{ B} \rangle &=& f(t) |0,t \rangle + g(t)
 \a_{-1}\tilde{\a}_{-1}|0,t\rangle + \cdots.
\eeq
By some algebra we have \cite{sen1}
\beq \label{scalarf} 
f(t) = \sum_j (-\pi g e^t)^{2j}  = \frac{1}{1+ \pi g e^t },\quad 
g(t) = f(t) - 2 = \frac{1}{1+ \pi g e^t} -2  
\eeq 
where the reverse Wick rotation is taken.

As $t \rightarrow \infty$, we find 
\beq
f(t) \rightarrow 0, \quad g(t) \rightarrow -2.
\eeq
Since the stress tensor for the tachyon profile is given as 
\beq
T_{00} = - T_p, \quad T_{ij} = \delta_{ij} T_p f(t),
\eeq
$T_{ij} \rightarrow 0 $ while $T_{00}$ is independent of time 
as $t \rightarrow \infty$. It implies that the pressure vanishes 
in this limit. This observation led Sen \cite{sen3,sen4} to propose that the decay 
product may be pressureless tachyon matter \cite{sugimo,buchel,ishida,kwon,yee}.

One may attempt to evaluate the higher level components in the 
expansion of the boundary state Eq.(\ref{expansion}). 
Constable and Larsen \cite{constable} developed an efficient techniques to evaluate 
the higher level components using the Polyakov string path integral. They 
compute large classes of the higher level components explicitly, 
in particular,
\beq
B^{(N,N)}(t) &=& f(t) - \frac{2}{N} \sum^{N-1}_{n=0} (N-n)(-\pi g 
e^t)^n,
\eeq
where $B^{(N,N)}$ is the coefficient of the terms $\alpha_{_N} 
\tilde{\alpha}_{-N} |0\rangle$ in the expansion of the boundary 
state Eq.(\ref{expansion}). An interesting point is that all the 
coefficients, $B^{(N,N)}$, diverge 
\beq
B^{(N,N)} \rightarrow -\frac{2}{N} (-\pi g)^{(N-1)} e^{(N-1)t} 
\qquad {\rm as} \,\,\, t \rightarrow \infty
\eeq
except for $B^{(1,1)}(t) = g(t) \rightarrow -2$. However, this divergent 
behaviour of $B^{(N,N)}(t)$ contrasts with our expectation: The 
boundary state $|{ B} \rangle$ is supposed to describe the time
dependent process of the rolling tachyon at classical level in a 
well defined manner. It leads us to speculate that the bosonic 
oscillator basis may not be a suitable basis to describe the time dependent 
description of the rolling tachyon.
  
The boundary condition which follows from the world-sheet string action 
Eq.(\ref{tachyonaction}) is
\beq
\left(\frac{1}{2\pi} \frac{\p~}{\p \t} X(\t,\s) - \frac{g}{2} 
e^X\right)\Biggl\vert_{\t=0} = 0.
\eeq
Note that when $g \rightarrow -\infty $, the boundary condition reduces to the 
Neumann condition. If we denote the zero mode part of $X$ by $t$, the 
effective coupling becomes $g e^t$ and the boundary condition 
implies that the unstable object is a D-brane initially in the far past. 
As $t \rightarrow \infty$, the boundary interaction becomes effectively stronger. And  
it has been conjectured \cite{gutperle02} that in the far future the unstable D-brane  
may become a S-brane \cite{strominger02,leblond03,chen02,kruc,roy02,deger,ohta02,wang,
iva,hashi,hashi03,jones}, 
which is a localized object in the temporal 
direction. The main subject of this paper is to show how 
one can realize this intuitive picture in the framework of the boundary 
state formulation, which takes into account the full string degrees of 
freedom.

\section{Time Evolution of the Boundary State}

One of the distinctive features of string theory which differentiates it 
from the ordinary quantum field theory is that ``time" is 
embedded in the dynamical variable $X$ along the temporal direction
as its zero mode. Due to 
this distinctive feature often time-dependent description
become subtle in string theory. In order to get the time-dependent 
description of a given dynamical process we should 
carefully factor the time from the dynamical degrees of freedom

Here we take a simple strategy to split the string variable $X$ as follows
\beq 
X = t + \hat X \label{split}.
\eeq
Here $t$ corresponds to the classical part of $X$, which we may call 
the "time" and $\hat X$ denotes collevtively the quantum degrees of freedom. 
We regard $t$ as a modular parameter of the string state (or the target space) and 
$\hat X$ as the dynamical variable describing quantum fluctuations around it. 
In order to have a well-define quantum system we may take the Wick rotation 
for $\hat X$ but not for $t$; 
$\hat X \rightarrow i \hat X$.\\

\begin{centerline}
{\bf Time Evolution of Some Simple Boundary States in One Dimension}
\end{centerline}

Recently we developed \cite{lee,hassel} a free 
fermion representation of the boundary conformal field theory for the rolling
tachyon, generalizing the work of Polchinski and 
Thorlacius \cite{Polchinski:1994my} for the open string to the closed string. 
In this fermion theory the boundary interaction becomes a simple 
fermion current operator. As a result an explicit, compact, 
exact expression of the boundary state has been obtained. Since the marginal boundary term becomes a 
bilinear operator in terms of fermion fields, it would be more 
appropriate to discuss the time dependent description using the 
fermion basis rather than the bosonic oscillator basis. 

Once we factor the time, we find that the fermion boundary state 
behaves distinctively depending on the boundary condition it 
satisfies. Let us consider a simple one dimensional system first where 
it has only one string coordinate. In two dimensions fermions and bosons are 
mapped to each other by
 \begin{subequations}
 \label{generallabel}
 \begin{eqnarray}
 \psi_L(z)~=~e^{-i\frac{\pi}{2}p_R}~:e^{-i\sqrt{2}X_L(z)}:
 ,~~~
 \psi_L^{\dagger}(z)~=~e^{i\frac{\pi}{2}p_R}~:e^{i\sqrt{2}X_L(z)}:
 \label{bosform:a}\\
 \psi_R(\bar z)~=~e^{-i\frac{\pi}{2}p_L}~:e^{i\sqrt{2}X_R(\bar
 z)}: ,~~~ \psi_R^{\dagger}(\bar
 z)~=~e^{i\frac{\pi}{2}p_L}~:e^{-i\sqrt{2}X_R(\bar z)}:
 \label{bosform:b}
 \end{eqnarray}
 \end{subequations}
 where  $X_R(\bar z)$ and $X_L(z)$ are the right- and
 left-moving boson fields, respectively.  
The left- and right-moving boson
operators are defined by the mode expansions:
 \begin{subequations}
 \label{generallabel}
 \begin{eqnarray}
 X_L(\tau+i\sigma)~=~
 \frac{1}{\sqrt{2}} x_L-\frac{i}{\sqrt{2}} p_L(\tau+i\sigma)+\frac{i}{\sqrt{2}}
 \sum_{n\neq 0}\frac{\a_n}{n}e^{-n(\tau+i\sigma)}, \label{expan:a} \\
 X_R(\tau-i\sigma)~=~
 \frac{1}{\sqrt{2}} x_R- \frac{i}{\sqrt{2}} p_R(\tau-i\sigma)+\frac{i}{\sqrt{2}}
 \sum_{n\neq 0}\frac{\tilde\a_n}{n}e^{-n(\tau-i\sigma)}. \label{expan:b}
 \end{eqnarray}
 \end{subequations}
The non-vanishing commutators are
 \begin{subequations}
 \label{generallabel}
 \begin{eqnarray}
 \left[ x_L,p_L\right]~&=&~i ~~~,~~~ \left[ x_R, p_R\right] ~=~ i
 \label{commu:a}\\
 \left[ \a_m, \a_n \right] ~&=&~ m\delta_{m+n} ~~~,~~~ \left[
 \tilde\a_m, \tilde\a_n \right] ~=~ m\delta_{m+n}.
 \label{commu:b}
 \end{eqnarray}
 \end{subequations}

The Neumann and Dirichlet boundary conditions for the bosonic string 
are given as 
 \begin{subequations}
 \label{generallabel}
 \begin{eqnarray}
 X_L(0,\sigma)\left|N \right\rangle&=& X_R(0,\sigma)\left|
 N \right\rangle, \label{LR:a}\\
 X_L(0,\sigma)\left| D \right\rangle&=&- X_R(0,\sigma)
 \left|D \right\rangle. \label{LR:b} \end{eqnarray}
 \end{subequations}
These boundary conditions are transcribed in the fermion theory 
as follows 
\begin{subequations}
\label{generallabel}
\begin{eqnarray}
 \psi_L(0,\sigma)~\left|N \right\rangle &=&
 i\psi^{\dagger}_R(0,\sigma) \left|N \right\rangle~,~~
 \psi_L^{\dagger}(0,\sigma) \left|N \right\rangle=
 i \psi_R(0,\sigma)~\left| N\right\rangle, \\
 \psi_L(0,\sigma)~\left| D\right\rangle &=&
 - i\psi_R(0,\sigma) \left| D\right\rangle~,~~
 \psi_L^{\dagger}(0,\sigma) \left| D\right\rangle=
 - i \psi_R^{\dagger}(0,\sigma)~\left| D\right\rangle.
\end{eqnarray}
\end{subequations}
We may construct the boundary states which satisfy these 
conditions\footnote{As is well-known, the fermion theory has two 
secotrs; the NS-NS sector and the R-R sector. Explicit 
expressions of the fermion boundary states are slightly 
different in each sector. Here we only discuss the NS-NS sector 
for the sake of simplicity. See \cite{lee} for more detailed expressions.}
\begin{subequations}
\label{generallabel}
\beq
 |N\rangle &=& :\exp\left\{i\int\frac{d\s}{2\pi} \left(\psi^\dagger 
 \tpsi^\dagger+ \psi\tpsi\right)\right\}:|0\rangle, \label{nbdr}\\
 |D\rangle &=& :\exp\left\{i\int\frac{d\s}{2\pi} 
 \left(\tpsi^\dagger\psi-\psi^\dagger \tpsi\right)\right\}:|0\rangle. 
 \label{dbdr}
\eeq
\end{subequations} 

If we introduce the time $t$ explicitly, the Neumann condition 
does not change but the Dirichlet condition changes into 
\beq
\left(X_L + X_R\right)|D;t\rangle = it |D;t\rangle.
\eeq 
The corresponding fermion boundary condition becomes
\begin{eqnarray} \label{fermit}
  \psi_L(0,\sigma)~\left| D; t\right\rangle &=&
 - i e^{\sqrt{2} t} \psi_R(0,\sigma) \left| D; 
 t\right\rangle~,\nn\\
 \psi_L^{\dagger}(0,\sigma) \left| D; t\right\rangle &=&
 - i e^{-\sqrt{2} t} \psi_R^{\dagger}(0,\sigma)~\left| D; t\right\rangle.
\end{eqnarray}
Hence the boundary state $|D;t\rangle$ is constructed to be in the fermion 
representation as
\beq
|D;t\rangle = :\exp\left\{i\int\frac{d\s}{2\pi}
(e^{-\sqrt{2}t} \tpsi^\dagger \psi - e^{\sqrt{2}t} \psi^\dagger \tpsi) 
\right\}:|0\rangle.
\eeq
Since the Neumann state does not change,
\beq
|N; t \rangle = |N; 0 \rangle = :\exp\left\{i\int\frac{d\s}{2\pi} \left(\psi^\dagger 
\tpsi^\dagger+ \psi\tpsi\right)\right\}:|0\rangle.
\eeq 

\newpage

\begin{centerline}
{\bf Time Evolution of Some Simple Boundary States in Two Dimensions}
\end{centerline}
In order to fermionize the boundary conformal 
field theory for the rolling tachyon 
we introduce an auxiliary free boson $Y$ as in 
refs. \cite{lee,hassel}
\begin{subequations}
 \label{generallable}
 \begin{eqnarray}\label{phi1}
 \phi_{1L}=\frac{X_L+Y_L}{\sqrt{2}}
 ~~,~~
 \phi_{1R}=\frac{X_R+Y_R}{\sqrt{2}}
 \\ \label{phi2}
 \phi_{2L}=\frac{X_L-Y_L}{\sqrt{2}}
 ~~,~~
 \phi_{2R}=\frac{X_R-Y_R}{\sqrt{2}}.
 \end{eqnarray}
 \end{subequations}
In the two dimensional system, described by two bosons
$X$ and $Y$, we have four simple boundary states $|D, N\rangle, \,\,
|D, D\rangle, \,\, |N, N\rangle, \,\, |N, D\rangle$ where the first and second 
label is the boundary condition for $X$ and $Y$ bosons respectively.
These simple boundary states are given in terms of the fermion 
fields respectively \cite{lee} as 
\beq
|N,N\rangle &=& :\exp\left\{i\int \frac{d\s}{2\pi} \left(\psi_1 
\tpsi_1+ \psi^\dagger_1 \tpsi^\dagger_1 + \psi_2 \tpsi_2 + 
\psi^\dagger_2 \tpsi^\dagger_2\right)\right\}:|0\rangle \nn\\
|N,D\rangle &=& :\exp\left\{\int \frac{d\s}{2\pi} \left(\psi^\dagger_1 
\tpsi_2- \psi^\dagger_2 \tpsi_1 - \tpsi^\dagger_1 \psi_2 + 
\tpsi^\dagger_2 \psi_1\right)\right\}:|0\rangle \\
|D,N\rangle &=&  :\exp\left\{\int \frac{d\s}{2\pi} \left(\psi^\dagger_1 
\tpsi^\dagger_2- \tpsi^\dagger_1 \psi^\dagger_2 + \tpsi_1 \psi_2 - 
\psi_1 \tpsi_2\right)\right\}:|0\rangle \nn\\
|D,D\rangle &=& :\exp\left\{i\int \frac{d\s}{2\pi} \left(\psi^\dagger_1 
\tpsi_1- \psi^\dagger_2 \tpsi_2 - \tpsi^\dagger_1 \psi_1 + 
\tpsi^\dagger_2 \psi_2\right)\right\}:|0\rangle \nn
\eeq 

Now we introduce $t$ by factoring the zero mode of $X$ explicitly. 
Since $|N,D\rangle$ and $|N,N\rangle$ are independent of time 
(the zero mode of $X$), they do not change in 
time
\beq
|N,N; t \rangle = |N,N; 0 \rangle = |N,N\rangle, \quad 
|N,D; t \rangle = |N,D; 0 \rangle = |N,D\rangle.
\eeq
The time dependence of the boundary state $|D,N\rangle$ follow from 
its $t$ dependent boundary condition
\beq
\psi_{1L}(0,\sigma) |D, N;t\rangle &=&
e^t \psi^{\dagger}_{2R}(0,\sigma)|D,N;t\rangle,
\quad \psi_{2L}(0,\sigma) |D, N\rangle = e^{-t}\psi^{\dagger}_{1R}(0,\sigma) |D,N;t\rangle, \\
\psi^{\dagger}_{1L}(0,\sigma) |D, N;t\rangle &=& -
e^{-t}\psi_{2R}(0,\sigma)|D,N\rangle, \quad \psi^{\dagger}_{2L}(0,\sigma)
|D, N\rangle = - e^{t}\psi_{1R}(0,\sigma) |D,N;t\rangle. \nn \eeq
Thus, the boundary state $|D,N;t\rangle$ is constructed as 
\begin{eqnarray}
|D,N;t\rangle= :\exp\left\{\int\frac{d\s}{2\pi}\left(
e^t \psi^\dagger_1 \tpsi^\dagger_2
-e^{-t} \tpsi^\dagger_1 \psi^\dagger_2  + 
e^{-t} \tpsi_1 \psi_2 - e^{-t}\psi_1 \tpsi_2\right) \right\}:|0\rangle.
\end{eqnarray}
One can construct the boundary state $|D,D;t\rangle$ also in a similar 
way. If $t$ is explicitly factored, the boundary condition for 
$|D,D;t\rangle$ is read as 
\begin{subequations}
\label{generallabel}
\beq
\psi_{1L}(0,\sigma) |D,D;t \rangle &=& i e^t \psi_{1R}(0,\sigma)|D,D;t\rangle, \quad
\psi_{2L}(0,\sigma)|D,D;t\rangle = -i e^{-t} \psi_{2R}(0,\sigma)|D,D;t\rangle, \\
\psi^{\dagger}_{1L}(0,\sigma)|D,D;t\rangle &=&
i e^{-t} \psi^{\dagger}_{1R}(0,\sigma)|D,D;t\rangle, \quad
\psi^{\dagger}_{2L}(0,\sigma) |D,D;t\rangle = -i e^t
\psi^{\dagger}_{2R}(0,\sigma)|D,D;t\rangle.  \eeq
\end{subequations} 
The boundary state $|D,D;t\rangle$ satisfying these boundary conditions is 
easily obtained as 
\beq \label{ddt}
&&|D,D;t\rangle =: \exp\left\{i\int \frac{d\s}{2\pi}\left(e^t 
\psi^\dagger_1 \tpsi_1 -e^{-t} \psi^\dagger_2 \tpsi_2 - 
e^{-t} \tpsi^\dagger_1 \psi_1 + e^t \tpsi^\dagger_2 
\psi_2\right)\right\}:|0\rangle.
\eeq

\begin{centerline}
{\bf Time Evolution of the Rolling Tachyon}
\end{centerline}

We may recall the exact boundary state for the rolling tachyon in 
fermion theory \cite{lee}
\beq 
|B,D; 0\rangle 
&=& :\exp\Biggl\{\int \frac{d\s}{2\pi} \left[\left(
\psi^\dagger_{1}\tpsi_{2} - \psi^\dagger_{2} \tpsi_{1}
- \tpsi^\dagger_{1}\psi_{2} + \tpsi^\dagger_{2} \psi_{1}\right)
- g\pi i  \left(\psi^\dagger_{1} \tpsi_{1}
-\tpsi^\dagger_{2} \psi_{2} \right)\right]\Biggr\}:|0\rangle 
\label{bdt} 
\eeq
The time dependent boundary condition $|B,D;t\rangle$ can be 
obtained by simply factoring $t$ 
\beq
|B,D; t\rangle 
&=& :\exp\Biggl\{\int \frac{d\s}{2\pi} \left[\left(
\psi^\dagger_{1}\tpsi_{2} - \psi^\dagger_{2} \tpsi_{1}
- \tpsi^\dagger_{1}\psi_{2} + \tpsi^\dagger_{2} \psi_{1}\right)
- g e^t\pi i  \left(\psi^\dagger_{1} \tpsi_{1}
-\tpsi^\dagger_{2} \psi_{2} \right)\right]\Biggr\}:|0\rangle 
\label{bdt2} 
\eeq
The boundary state $|B,D;t\rangle$ may be considered as the quantum 
state of the unstable D-brane probed by the closed string around 
$\langle X \rangle = t$. 

\section{The Final Fate of the Rolling Tachyon}

We may rewrite the boundary state $|B,D; t\rangle $ Eq.(\ref{bdt2}) as follows 
\beq \label{bdt3}
|B,D; t\rangle &=& 
:\exp\left\{i \int \frac{d\s}{2\pi} \left[\Psi^\dagger_L
M_1 \Psi_R
-\Psi^\dagger_R M_2 \Psi_L\right] 
\right \}:|0\rangle 
\eeq
where
\beq
M_1 = \s_2 + \frac{ge^t\pi}{2} (I + \s_3) = \left(\begin{array}{cc}
  ge^t\pi & -i \\
  i & 0 
\end{array}\right), \quad 
M_2 = \s_2 + \frac{ge^t\pi}{2} (I - \s_3) = \left(\begin{array}{cc}
  0 & -i \\
  i & ge^t\pi 
\end{array}\right) 
\eeq
and 
\beq
\Psi_L = \left(\begin{array}{c}
  \psi_1 \\
  \psi_2 
\end{array}\right),\quad 
\Psi_R = \left(\begin{array}{c}
  \tpsi_1 \\
  \tpsi_2 
\end{array}\right). \nn
\eeq
Note that $\Psi_L$ and $\Psi_R$ 
are the local perturbative basis in string theory and we are 
free to choose the most suitable one to describe the system. 
Any two basis are related
to each other by a similarity transformation. 
The analysis of the time dependent behaviour in the limit 
where $t \rightarrow \infty$ would be simpler if we choose 
the perturbative basis where the matrices $M_1$ and $M_2$ are 
diagonalized. What we should be concerned 
with is the spectrum of the string state in the limit. 

Both matrices $M_1$ and $M_2$ are Hermitian 
and have the same characteristic equation: 
Their eigenvalues are
\beq
\lambda_\pm = \frac{ge^t\pi}{2} \pm \sqrt{1+ 
\left(\frac{ge^t \pi}{2}\right)^2}.
\eeq
The (normalized) eigenvectors for $M_1$ with the eigenvalues 
$\lambda_\pm$ respectively are  
\beq
\frac{1}{\sqrt{\lambda^2_\pm +1}}\left(\begin{array}{c}
  \lambda_\pm \\
  i
\end{array}\right).
\eeq
Defining
\beq
S_1 = \left(\begin{array}{cc}
  \frac{\lambda_+}{\sqrt{\lambda^2_++1}} & \frac{\lambda_-}{\sqrt{\lambda^2_-+1}} \\
  \frac{i}{\sqrt{\lambda^2_++1}} & \frac{i}{\sqrt{\lambda^2_-+1}} 
\end{array}\right), 
\eeq
we diagonalize $M_1$ by a similarity transformation
\beq
M_1 \rightarrow S^{-1}_1 M_1 S_1 = \left(\begin{array}{cc}
  \lambda_+ & 0 \\
  0 & \lambda_- 
\end{array} \right).
\eeq
Simultaneously we transform the fermion fields $\Psi^\dagger_{-n}$ 
and $\tPsi_{-n}$ with $n>0$
\beq
\Psi^\dagger_{-n} \rightarrow \Psi^\dagger_{-n} S_1, \quad
\tPsi_{-n} \rightarrow S^{-1}_1 \tPsi_{-n}.
\eeq
In order to preserve the canonical anti-commutation relations among 
fermion fields we must also transform $\Psi_n$ and $\tPsi^\dagger_n$ 
($n>0$) as follows  
\beq
\Psi_n \rightarrow S^{-1}_1 \Psi_n, \quad 
\tPsi^\dagger_n \rightarrow  \tPsi^\dagger_n S_1.
\eeq

Following the same steps we diagonalize $M_2$. If we choose the 
normalized eigenvectors for $M_2$ as 
\beq
\frac{1}{\sqrt{\lambda^2_\mp +1}}\left(\begin{array}{c}
 -i\\
  \lambda_\mp
\end{array}\right),
\eeq
the transformation matrix $S_2$ is given as 
\beq
S_2 = \left(\begin{array}{cc}
  - \frac{i}{\sqrt{\lambda^2_- +1}} & - \frac{i}{\sqrt{\lambda^2_++1}} \\
  \frac{\lambda_-}{\sqrt{\lambda^2_-+1}} & \frac{\lambda_+}{\sqrt{\lambda^2_++1}} 
\end{array} \right).
\eeq
The second term in the exponent of Eq.(\ref{bdt2}) can be diagonalized 
by taking a similarity transformation
\beq
M_2 \rightarrow - S^{-1}_2 M_2 S_2, \quad 
\Psi_{-n} \rightarrow i S^{-1}_2 \Psi_{-n}, \quad 
\tPsi^\dagger_{-n} \rightarrow i \tPsi^\dagger_{-n} S_2, \quad n>0
\eeq
Here an extra phase $i$ is introduced to take care of the unwanted 
phase we will get after diagonalization.
The canonical anti-commutation relation is preserved if we transform 
$\Psi^\dagger_n$ and $\tPsi_n$ accordingly as 
\beq
\Psi^\dagger_n \rightarrow -i \Psi^\dagger_n S_2, \quad 
\tPsi_n \rightarrow -i S^{-1}_2 \tPsi_n, \quad n>0
\eeq
Upon diagonalizing, we have
\begin{subequations}
\label{generallabel}
\beq
M_1 &=&  \frac{g\pi}{2} I + \sqrt{1+ \left(\frac{g e^t \pi}{2}\right)^2} \s_3,
\\
M_2 &=& - \frac{g\pi}{2} I + \sqrt{1+ \left(\frac{g e^t\pi}{2}\right)^2} \s_3. 
\eeq
\end{subequations}
(Note also that under this similarity transformation the world sheet string
Hamiltonian is invariant.)
 
In the limit where $t \rightarrow \infty$,
\beq
M_1 \rightarrow  \left(\begin{array}{cc}
  e^t & 0 \\
  0 & -e^{-t} 
\end{array} \right), \quad
M_2 \rightarrow \left(\begin{array}{cc}
  e^{-t} & 0 \\
  0 & - e^{t} 
\end{array} \right).
\eeq
Here we translate $t$ such that
\beq
t \rightarrow t+ t_0, \quad g \pi e^{t_0} = 1.
\eeq
After diagonalizing the exponent, in the limit where $t \rightarrow 
\infty$, the boundary state $|B,D;t\rangle$ behaves precisely as 
$|D,D;t\rangle$ Eq.(\ref{ddt})
\beq
|B,D;t\rangle \rightarrow |D,D;t\rangle,\quad 
{\rm as}~~t \rightarrow \infty
\eeq
Of course, in the far past where $t \rightarrow -\infty$,
the boundary state reduces to the boundary state for a 
D-brane
\beq
|B,D;t\rangle \rightarrow |N,D;t\rangle = |N,D\rangle, \quad 
{\rm as}~~t \rightarrow -\infty
\eeq
Thus, in the far future the unstable Dp-brane can be 
viewed as a Sp-brane if the local perturbative basis is 
appropriately chosen. 

\section{Conclusions}

The time dependent decay process of the unstable D-brane is one of 
the most important subjects in string theory. Splitting the 
classical part from the quantum part of the string coordinate 
variable in the temporal direction and identifying it as time,
we observe the time dependent behaviour of the boundary state, 
which describes the unstable D-brane. The classical part called 
time is a non-dynamical part of the zero mode, which
may be regarded as a modular parameter of the target space-time. 
The boundary state proposed by Sen \cite{sen1} is supposed to be 
a classical solution of the string theory which should depict the decay process at 
the classical level. The final fate of the unstable D-brane, which 
should be described by the proposed boundary state, has been one of focal points 
of recent studies \cite{senreview,lambert,sen022,mukhopa,sen023,sen0305,sen031,sen0306,sen0308,sen04,
larsen02,gibbons,okuda,kim,hlee,rey,rey2,taka,doug,arefeva,demasure,gutperle03,larsen,
constable,schomerus,nagami,foto,coletti,forini}.
It has been also conjectured that the final decay product may be the S-brane, all of 
which tangential dimensions are spacelike \cite{gutperle02,strominger02,leblond03,chen02,kruc,roy02,deger,ohta02,wang,
iva,hashi,hashi03,jones}. 
However, how such a classical description of 
the decay process of the unstable D-brane can be realized in a consistent manner 
with the rolling tachyon was an open question. 

Here in this paper we propose an alternative interpretation of the 
boundary state for the rolling tachyon to show that the boundary state 
correctly depicts the decay process of the unstable D-brane into 
a S-brane at classical level. The strategy we take is to separate 
the non-dynamical part from the string coordinate variable in the 
temporal direction and identify it as time and the rest as the 
quantum degrees of freedom. Then we apply the Wick rotation to quantum 
part only to have a well-defined quantum system as suggested by Sen and  
examine the time dependent behaviour of the boundary state.
In the far past the boundary state trivially reduces to that for a D-brane.
And we observe that in the far future the boundary state approaches that of 
the S-brane if we choose the local string perturbative basis appropriately.
The boundary state delineates continuous transition of the D-brane into a 
S-brane at classical level. It is worth while to note that the 
fermion representation of the boundary state is quite useful to 
find the most suitable perturbative basis. 

The unstable D-brane may undergo further some quantum decay 
process by emitting closed strings. The quantum decay process 
of the unstable D-brane has been already studied by numerous authors
\cite{lambert,chen0209,mcgree,kleb,kluson,nagami03,hashi04,shelton,leblond,
jokela,ckim05}. It would be an interesting work to combine the classical 
process discussed here and the quantum process studied in the previous works
to get a consistent unified description of the decay of the unstable D-brane.

\section*{Acknowledgement}
Part of this work was during the author's visit to ICTP (Italy), KIAS (Korea) 
and PITP (Canada). This work was supported by KOSEF through Center 
for Quantum Space-Time (CQUeST).

\end{document}